\documentclass[a4paper,twocolumn,showpacs,preprintnumbers,aps,amsmath,amssymb,nofootinbib]{revtex4}
\input psfig.sty 
\usepackage{picinpar}
\usepackage[dvips]{graphicx}

\begin{document}

\title{Scalar field description of a parametric model of dark energy}

\author{E. M. Barboza Jr.$^{1,2}$\footnote{E-mail: edesiobarboza@uern.br}}

\author{J. S. Alcaniz$^2$\footnote{E-mail: alcaniz@on.br}}

\author{B. Santos$^2$\footnote{E-mail: thoven@on.br}}

\affiliation{$^1$Departamento de F\'isica, Universidade do Estado do Rio Grande do Norte, 59610-210, Mossor\'o - RN, Brasil}

\affiliation{$^2$Observat\'orio Nacional, 20921-400, Rio de Janeiro - RJ, Brasil}

\date{\today}

\begin{abstract}
We investigate theoretical and observational aspects of  a time-dependent parameterization for the dark energy equation of state (EoS) $w(z)$, which is a well behaved function of the redshift $z$ over the entire cosmological evolution, i.e., $z \in [-1,\infty)$. By using a theoretical algorithm of constructing the quintessence potential directly from the effective EoS parameter, we derive and discuss the general features of the resulting potential for this $w(z)$ function. Since the parameterization here discussed allows us to divide the parametric plane in defined regions associated to distinct classes of dark energy models, we use the most recent observations from type Ia supernovae, baryon acoustic oscillation peak and Cosmic Microwave Background shift parameter to check which class is observationally prefered. We show that the largest portion of the confidence contours lies into the region corresponding to a possible crossing of the so-called phanton divide line at some point of the cosmic evolution.

\end{abstract}


\maketitle

\section{Introduction}

Cosmological models with cold dark matter plus a $\Lambda$ term ($\Lambda$CDM) may explain most of the current astronomical observations (see, e.g., \cite{review} for some recent reviews). However, from the theoretical viewpoint it is really difficult to reconcile the small value required by observations ($\simeq 10^{-10} \rm{erg/cm^3}$) with estimates from quantum field theories ranging from 50-120 orders of magnitude larger, as well as to explain why this is exactly the right value that is just beginning to dominate the energy density of the Universe today.

These issues make a complete cancellation of $\Lambda$ (from some unknown symmetry of Nature) seem a plausible possibility and have also motivated a number of alternative explanations for the cosmic acceleration (for some of these alternative scenarios, see~\cite{other, mad}). One of these possibilities, possibly the next simplest approach toward an accelerating model for the Universe, is to work with the idea that the dark energy component is due to a minimally coupled scalar field $\phi$ which has not yet reached its ground state and whose current dynamics is basically determined by its potential energy $V(\phi)$~\cite{quint}. Clearly, however, such a procedure cannot provide a model-independent parameter space to be compared with the observational data.

Another way, widely explored in the literature,  is to build a phenomenological functional form for the dark energy equation of state (EoS), i.e., the ratio of its pressure to its energy density, $w \equiv p/\rho$, in terms of its current value $w_0$ and of its time-dependence $w_a =dw/d\ln a$, and study its cosmological consequences as well as possible constraints on its behavior from observations. Usually, these parameterizations have not only the standard $\Lambda$CDM  scenario ($w_0 = -1; w_a= 0$) but also the so-called $w$CDM model ($w_a = 0)$ as particular cases, so that constraints on its parameters may provide more accurate consistency checks to the original models.

Examples of  some EoS parameterizations recently explored are (see also~\cite{para,para1,generalized}):
\begin{eqnarray}
\label{OtherParameterizations}
w(z) = \; \left\{
\begin{tabular}{l}
$w_0 + w_a z$ 
\, 
\quad \quad  \quad  \quad \quad \quad \quad \quad  \hspace{0.35cm}\cite{linear}\\
\\
$w_0+w_az/(1+z)$ 
\quad \quad  \quad \quad \quad \quad \cite{cpl} \\
\\
$w_0-w_a\ln(1+z)$ 
\quad \quad  \quad \quad \quad \quad \cite{efs}
\end{tabular}
\right.
\end{eqnarray}
An interesting aspect worth mentioning is that it is difficult to obtain the above parameterizations from usual scalar field dynamics since they are not limited functions, i.e., the EoS parameter does not lie in the interval $w \in[-1,1]$. In other words, this amounts to saying that when extended to the entire history of the universe, $z\in [-1,\infty)$, the three parameterizations above are divergent functions of the redshift (see also \cite{para1} for a discussion)\footnote{The $w(z)$ parameterizations given by Eq. (1) can be expressed as a single and generalized EoS function, i.e.,  $w(z) = w_0 - w_a\frac{(1+z)^{-\beta} -1}{\beta}$, where the parameter $\beta$ takes, respectively, the values $\pm 1$ and $0$~\cite{generalized}.}.

In Ref.~\cite{edesio} we investigated some cosmological consequences of a new phenomenological parameterization:
\begin{equation}
\label{MyParameterization}
w(z)=w_0+w_a\frac{z(1+z)}{1+z^2}\;,
\end{equation}
or, equivalently,
\begin{equation}
\label{MyParameterization1}
w(z)=w_0+w_a\frac{1-a}{1 - 2a + 2a^2}\;,
\end{equation}
which has the same linear behavior at low redshifts presented by the parameterizations discussed above but with the advantage of being a limited function of $z$ throughout the entire history of the Universe (see also~\cite{pavon} for a recent analysis of a coupled quintessence model driven by a dark energy component parameterized by (\ref{MyParameterization})).

Our goal in this paper is twofold. First, to derive a scalar field description for the dark component whose EoS parameter is given by Eq. (\ref{MyParameterization}). To that end, we use the theoretical method of constructing the dark energy potential $V(\phi)$  directly from the effective EoS, as developed in Ref.~\cite{method}. We also generalize this algorithm to include the so-called phantom case. Second, to place constraints on the ($w_0$, $w_a$) plane of parameterization (\ref{MyParameterization}) from current observational data to check which class of dark energy is observationally prefered. We use some of the most recent type Ia supernovae (SNe Ia) observations, namely, the so-called Union2 sample of 557 events~\cite{union}, the  nearby + SDSS + ESSENCE + SNLS + Hubble Space Telescope (HST) set of 288 SNe Ia discussed in Ref.~\cite{sdss} (which we refer to as SDSS compilation). We consider two sub-samples of this latter compilation that use SALT2~\cite{salt2} and MLCS2k2~\cite{mlcs2k2} SN Ia light-curve fitting method. Along with the SNe Ia data, and  to help break the degeneracy between the dark energy parameters we also use  measurements of the baryonic acoustic oscillation (BAO) peak at $z = 0.2, 0.35$ and 0.6~\cite{bao,percival,blake} and the current estimate of the CMB shift parameter of Ref.~\cite{wmap}.

\begin{figure*}[t]
\centerline{\psfig{figure=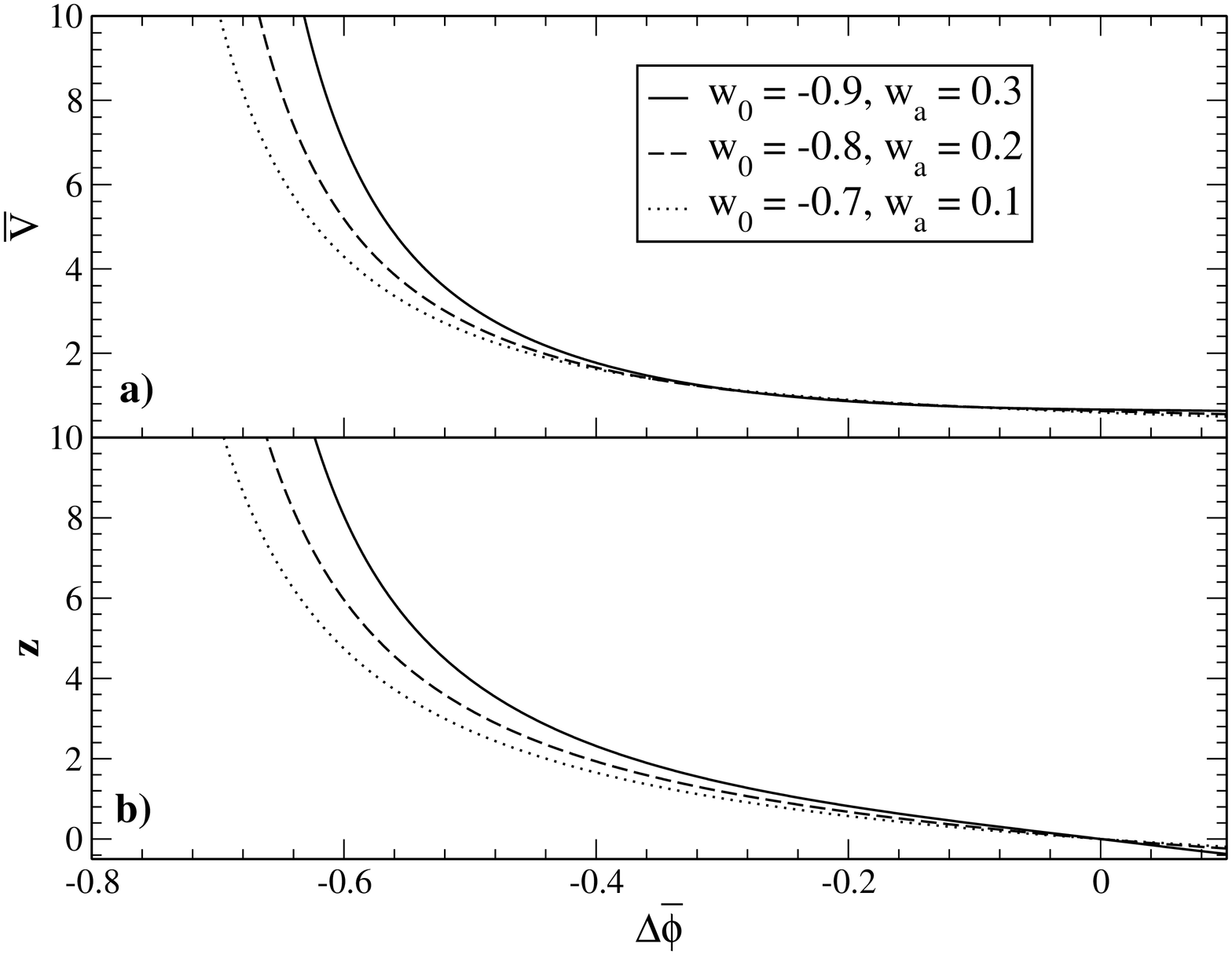,width=3.4truein,height=3.0truein,angle=270}
\hspace{0.5cm}
\psfig{figure=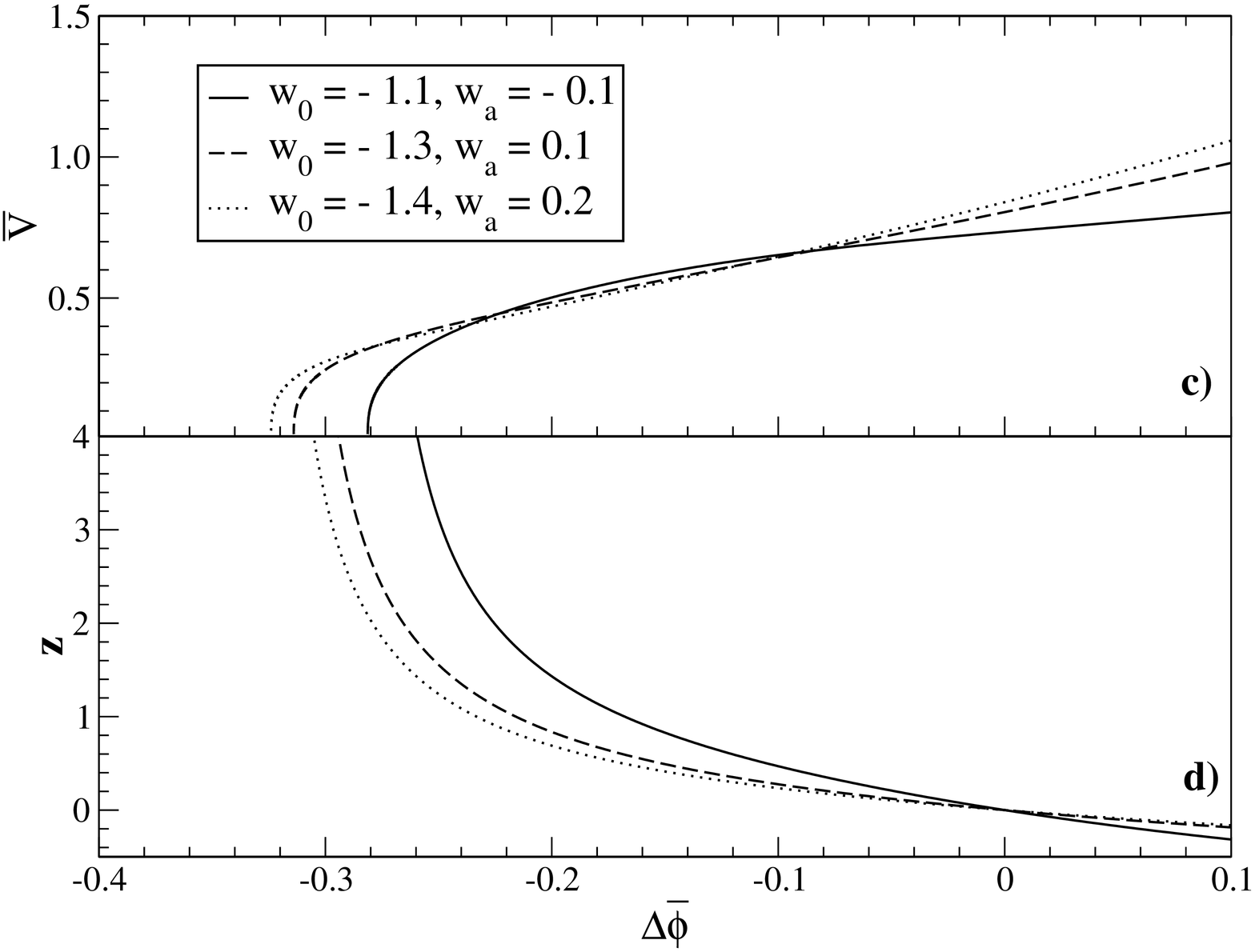,width=3.4truein,height=3.0truein,angle=270}
\hskip 0.1in}
\caption{Scalar field description of parameterization (\ref{MyParameterization}) for three selected points in the quintessence (Panels a and b) and phantom (Panels c and d) regions.}
\end{figure*}


\section{Scalar field description}

Let us first consider a homogeneous, isotropic, spatially flat cosmologies described by the Friedmann-Robertson-Walker (FRW) flat line element, $ds^2 = dt^2 - a^2(t)(dx^2 + dy^2 + dz^2)$, where we have set the speed of light $c = 1$. In such a background, the Friedmann equation for a dark matter-dark energy dominated universe is given by
\begin{equation}
H^2 = \frac{8\pi G}{3}(\rho_m + \rho_{\phi})\;,
\end{equation}
where $\rho_m$ stands for the non-relativistic (dark plus baryonic) matter. The energy density and pressure of the dark energy field are given by
\begin{subequations}
\begin{equation} \label{rhophi}
\rho_{\phi} = \epsilon\frac{1}{2}\phi^2 + V(\phi)\;,
\end{equation}
\begin{equation} \label{pphi}
p_{\phi} = \epsilon\frac{1}{2}\phi^2 - V(\phi)\;,
\end{equation}
\end{subequations}
where $\epsilon=\pm 1$ stands for quintessence ($-1 < w_{\phi} \leq -1/3$)~\cite{quint} and phantom fields ($w_{\phi} < -1$)~\cite{phantom}, respectively\footnote{Here, we generalize the results of \cite{method} and consider the possibility of phantom fields.}.

By combining Eqs. (\ref{rhophi}) and (\ref{pphi}), we obtain
\begin{subequations}
\begin{equation}
\label{dphi}
\dot{\phi}^2=\frac{1+w_{\phi}}{\epsilon}\rho_{\phi}\;,
\end{equation}
and
\begin{equation}
\label{vphi}
V(\phi)=\frac{1}{2}(1-w_{\phi})\rho_{\phi}\;,
\end{equation}
or still, in terms of $z$,
\begin{equation}
\dot{\phi}=\frac{d\phi}{dz}\dot{z}=-\frac{d\phi}{dz}(1+z)H(z)\;,
\end{equation}
so that,
\begin{equation}
\frac{d\phi}{dz}=\pm\frac{1}{(1+z)H(z)}\sqrt{\frac{1+w_{\phi}}{\epsilon}\rho_{\phi}}\;,
\end{equation}
\end{subequations}
where the negative (positive) signs stands to $\dot{\phi}>0$ ($\dot{\phi}<0$). Here, we adopt the negative sign.

By defining $\bar{\phi}\equiv\sqrt{8\pi\,G/3}\phi$ and $\bar{V}\equiv V/\rho_{c,0}$ and taking into account that $(1+w_{\phi})/\epsilon=\vert1+w_{\phi}\vert$, we have
\begin{subequations}
\begin{eqnarray}
\label{phitilde}
\Delta\bar{\phi}&\equiv&\bar{\phi}-\bar{\phi}_0 \nonumber\\ &=&-\int_0^z\frac{1}{(1+z)\eta(z)}\sqrt{\vert1+w_{\phi}(z)\vert\Omega_{\phi,0}f(z)}
\end{eqnarray}
and
\begin{equation}
\label{vphitilde}
\bar{V}(\bar{\phi})=\frac{1}{2}[1-w_{\phi}(z)]\Omega_{\phi,0}f(z),
\end{equation}
\end{subequations}
where $\eta(z)=H(z)/H_0$ and $f(z)$ stands for the time-dependent part of the energy density of the dark component, which, for parameterization (\ref{MyParameterization}), is given by
\begin{eqnarray}
f(z) & = & \frac{\rho_\phi}{\rho_{\phi,0}} \nonumber\\ & = & (1+z)^{3(1+w_0)}(1+z^2)^{3w_a/2}\;.
\end{eqnarray}
Note also that Eqs. (\ref{phitilde}) and (\ref{vphitilde}) are valid for both quintessence and phantom fields. By combining numerically these equations and taking into account the above constraints, we show in Figures 1a and 1c the resulting potential $\bar{V}(\bar{\phi})$ for parameterization (\ref{MyParameterization}) for quintessence and phantom regimes, respectively. Figures 1b and 1d show the evolution of the dark energy field as function of the redshift. The selected points used to plot the curves belong to quintessence and phantom families with $w ({z >>1})= w_0+w_a = -0.6$ and $w({z >>1})= w_0+w_a=-1.2$ and follow the constraints given in Sec. III. We note that, in contrast with a canonic scalar field in which the potential increases with the redshift (Panel 1b), for the phantom case shown in Panel 1d $\bar{V}(\bar{\phi})$ decrease with $z$. This result can be more easily understood by considering Eq. (\ref{vphitilde}), i.e., for $\quad z\gg1$,  $w_{\phi}\to w_0+w^{\prime}_0 <-1$ {and} $f(z)\to z^{-3\vert1+w_0+w_a\vert}\to0$, so that for phantom fields $\bar{V}(\bar{\phi})\to0$ when $z\to\infty$.

\section{Constraints on the $w_0 - w_a$ plane}

Before proceeding to the observational analyses on the $w_0 - w_a$ parametric plane, it is worth mentioning that the parameters $w_0$ and $w_a$ of (\ref{MyParameterization}) are subject to the following constraints:
$$
-1\leq w_0-0.21w_a\ \  {\mbox{and}} \ \  w_0+1.21w_a\leq1 \ \ ({\mbox{if}} \ \ w_a>0)
$$
and
$$
-1\leq w_0+1.21w_a\ \  {\mbox{and}} \ \  w_0-0.21w_a\leq1 \ \ ({\mbox{if}} \ \ w_a<0)
$$
for a quintessence-like behavior and
$$
w_a<-(1+w_0)/1.21 \ \ ({\mbox{if}} \ \ w_a>0)
$$
and
$$
w_a>(1+w_0)/0.21  \ \ ({\mbox{if}} \ \ w_a<0)\;
$$
for phantom fields. These bounds allow us to divide the parametric plane ($w_0 - w_a$) in defined regions associated to distinct classes of dark energy models that can be confronted with current observational data. These regions are shown in the Figures 2a-2c, where the area of early dark energy dominance corresponds to the constraint $w_0+w_a<0$, required to ensure $\rho_m(z)>\rho_{\phi}(z)$ at $z\gg1$. The blank regions indicate models that at some point of the cosmic evolution, $z \in  [-1,\infty)$, have switched or will switch from quintessence to phantom behaviors or vice-versa.

\begin{figure*}
\centerline{
\psfig{figure=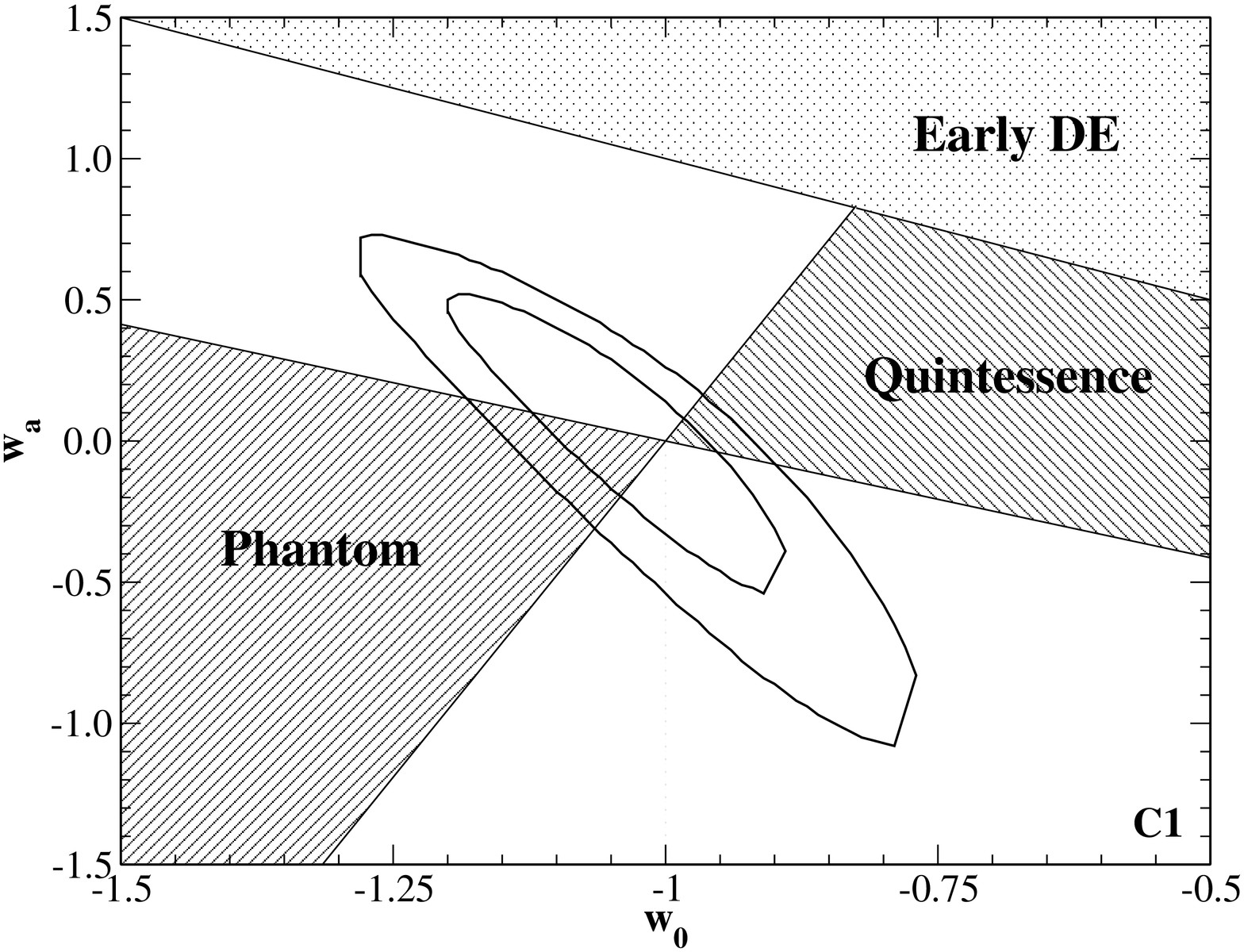,width=2.25truein,height=2.5truein,angle=270}
\psfig{figure=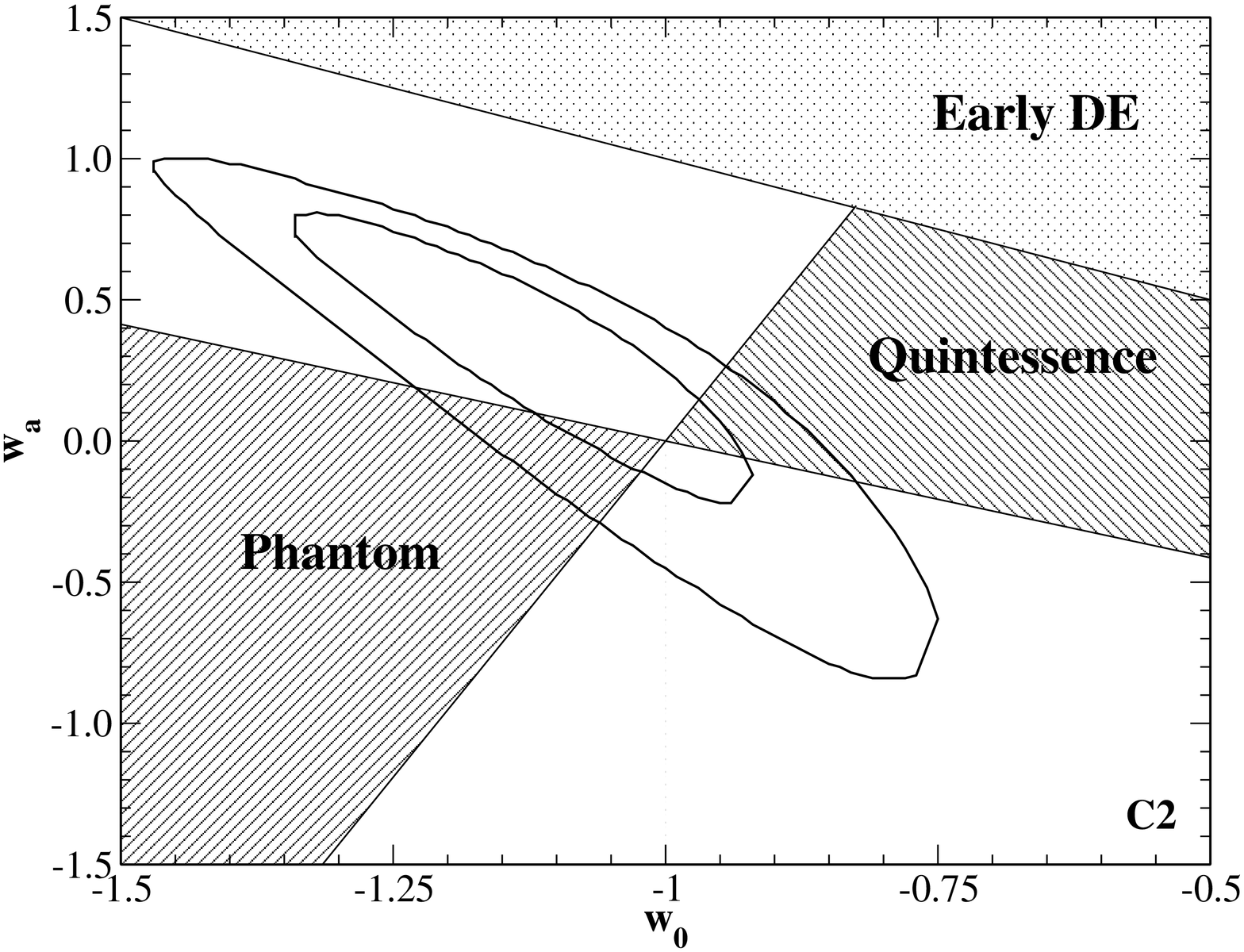,width=2.25truein,height=2.5truein,angle=270}
\psfig{figure=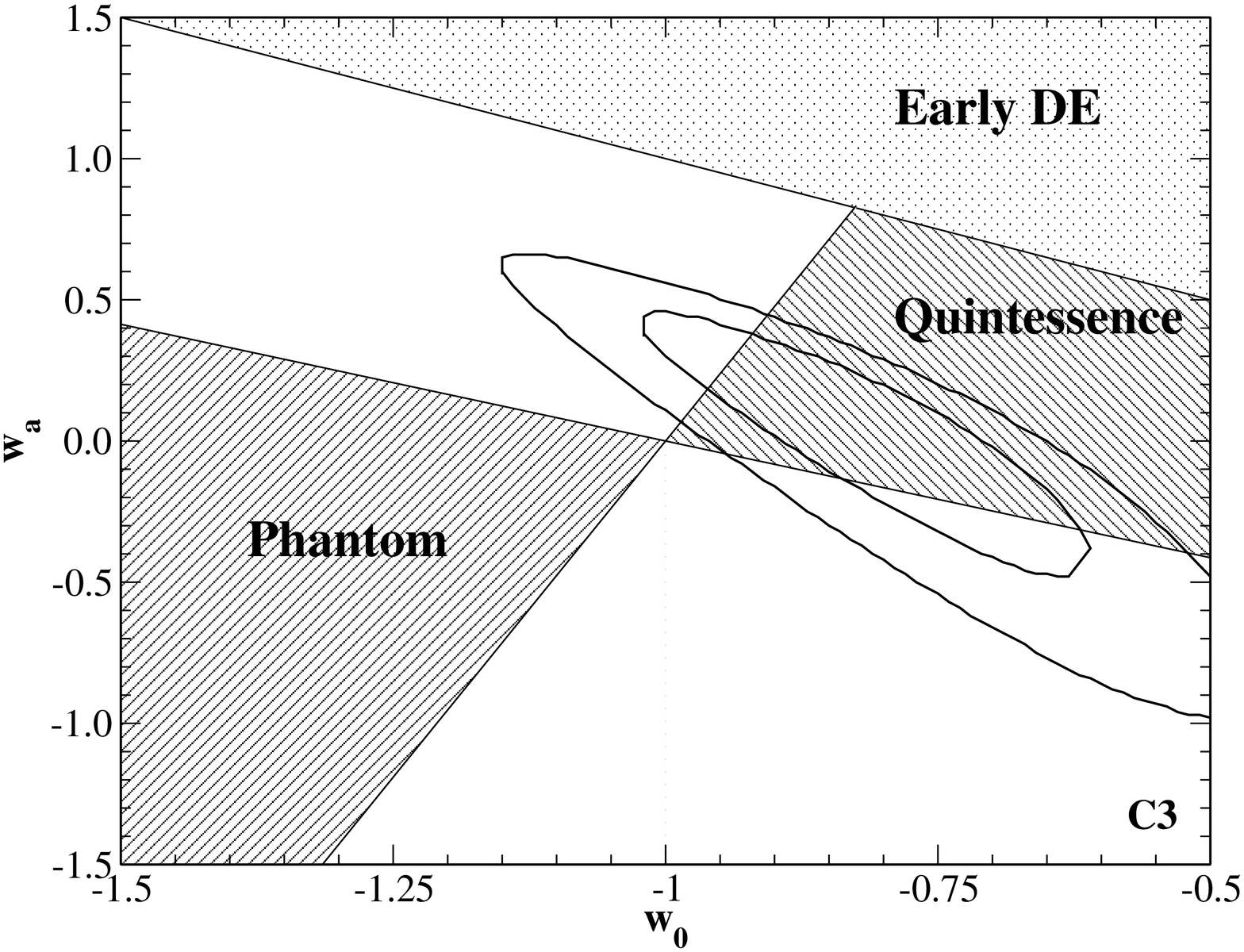,width=2.25truein,height=2.5truein,angle=270}}
\caption{The $w_0$ - $w_a$ parametric space for parameterization (2). As discussed in the text, the blank regions indicate models that at some point of the cosmic evolution have switched or will switch from quintessence to phantom behaviors or vice-versa. The Panels also show the resulting $\chi^2$ contours for the joint analyses C1, C2 and C3. The contours are drawn for $\Delta \chi^2 = 2.30$ and 6.17.}
\end{figure*}

\subsection{Observational constraints}

In order to discuss the current observational constraints on $w_0$ and $w_a$ we  use two of the most recent SNe Ia compilations available, namely, the Union2 sample of  Ref.~\cite{union} and the two compilations of the SDSS collaborations discussed in Ref.~\cite{sdss}. The Union2 sample is an update of the original Union compilation. It comprises 557 data points including recent large samples from other surveys and uses SALT2 for SN Ia light-curve fitting~\cite{union}. The SDSS compilation of 288 SNe Ia uses both SALT2 and MLCS2k2 light-curve fitters and is distributed in redshift interval $0.02 \leq z \leq 1.55$~\cite{sdss}.

Along with the SNe Ia data, and to diminish the degeneracy between the dark energy parameters $\Omega_{m,0}$, $w_0$ and $w_a$, we also use the results of current BAO and CMB experiments. For the BAO measurements, we use the three estimates of the BAO parameter ${\cal{A}} (z) = D_V{\sqrt{\Omega_{\rm{m}} H_0^2}}$ at $z = 0.2$, $0.35$ and $0.6$, as given in Table 2 of Ref.~\cite{blake}. In this latter expression, $D_V = [r^2(z_{\rm{BAO}}){z_{\rm{BAO}}}/{H(z_{\rm{BAO}})}]^{1/3}$ is the so-called dilation scale, defined  in terms of the dimensionless comoving distance $r$. For the CMB, we use only the
measurement of the CMB shift parameter~\cite{wmap}
\begin{equation}
{\cal{R}} = \Omega_{\rm{m}}^{1/2}r(z_{\rm{CMB}}) = 1.725 \pm 0.018\;,
\end{equation}
where $z_{\rm{CMB}} = 1089$. In our analyses, we minimize the function $\chi^2 = \chi^{2}_{\rm{SNe}} + \chi^{2}_{\rm{BAO}} + \chi^{2}_{\rm{CMB}}$, which takes into account all the data sets mentioned above and marginalize over the present value of the Hubble parameter $H_0$. For simplicity, the combinations Union2 + BAO + CMB, SDSS (SALT2) + BAO + CMB and SDSS (MLCS2k2) + BAO + CMB are denoted by C1, C2 and C3 respectively.

Panels 2a-2c show the results of our statistical analyses. In order to compare the theoretical frame with the observational constraints discussed above the three-dimensional parameter space $\Omega_{\rm{m}} - w_0 - w_a$ has been projected into the plane $w_0 - w_a$. Panels 2a, 2b and 2c show contours of $\Delta \chi^2 = 2.30$ and 6.17 arising from the joint analyses C1, C2 and C3, respectively. With exception of C3, we note that no dark energy behavior is preferred or ruled out by observations, although the largest portion of the confidence contours lies into the blank region, which indicates a possible crossing of the so-called phanton divide line at some point of the cosmic evolution (see \cite{divideline} for a discussion).

\begin{table}
	\begin{center}
		\begin{tabular}{ccccc}
		\hline
		\hline\\
		&$\chi^2_{min}$&\quad $w_0$&\quad $w_a$&\quad $\Omega_{m,0}$\\

		\hline
		\hline\\
		C1.............&545.02&$-1.05^{+0.22}_{-0.19}$&$0.08^{+0.54}_{-0.88}$&$0.28^{+0.02}_{-0.02}$\\
		C2.............&249.14&$-1.14^{+0.30}_{-0.27}$&$0.41^{+0.50}_{-0.93}$&$0.28^{+0.02}_{-0.02}$\\
		C3.............&242.09&$-0.82^{+0.29}_{-0.27}$&$0.06^{+0.50}_{-0.77}$&$0.31^{+0.02}_{-0.02}$\\
		\hline
		\hline
		\end{tabular}
	\end{center}
	\caption{The results of our analyses for C1, C2 and C3. The error bars correspond to 95.4\% C.L.}
	\label{bfz}
\end{table}

In Table I we summarize the results of our statistical analyses. For the sake of comparison, we also performed the C1, C2 and C3 analyses for the so-called CPL parameterization, $w(z) = w_0 + w_a z/(1+z)$~\cite{cpl}. We have found very similar values to those shown in Table I. This means that, although  different from the theoretical viewpoint -- CPL parameterization blows up exponentially in the future as $z \rightarrow -1$  for $w_a > 0$ whereas Parameterization (1) is a limited function of $z$ $\forall z\in [-1,\infty)$ -- both parameterizations provide very similar description for the current observational data.

\section{Conclusions}

We have examined theoretical and observational aspects of the EoS parameterization given by Eq. (\ref{MyParameterization})~\cite{edesio}. By following the method of constructing the quintessence potential directly from the effective equation of state function developed in Ref.~\cite{method}, we have derived the scalar field description for this $w(z)$ parameterization and extended our results for the case of phanton fields in which $w(z) < -1$. We also have performed a joint statistical analysis involving some of the most recent cosmological measurements of SNe Ia, BAO peak and CMB shift parameter. From a pure observational perspective, we have shown that both quintessence and phantom behaviours are acceptable regimes. In agreement with recent analyses, it has been shown that the largest portion of the confidence contours arising from these observations lies in the region of models that have crossed or will eventually cross the so-called phanton divide line at some point of the cosmic evolution.

\begin{acknowledgments}
This work is supported by Conselho Nacional de Desenvolvimento Cient\'ifico e Tecnol\'ogico (CNPq - Brazil).
\end{acknowledgments}


\begin{thebibliography}{99}

\bibitem{review} V.~Sahni and A.~A.~Starobinsky, Int.\ J.\ Mod.\ Phys.\  D {\bf 9}, 373 (2000); P. J. E. Peebles and B. Ratra Rev. Mod. Phys. {\bf{75}}, 559 (2003);  T. Padmanabhan, Phys. Rept. {\bf{380}}, 235 (2003); E. J. Copeland, M. Sami and S. Tsujikawa, Int. J. Mod. Phys. {\bf{D15}}, 1753 (2006); J. S. Alcaniz, Braz. J. Phys. {\bf{36}}, 1109 (2006). astro-ph/0608631; B. Ratra \& M.\ S.\ Vogeley, PASP  {\bf{120}}, 235 (2008);  A. Silvestri \& M. Trodden, Rept.\ Prog.\ Phys. {\bf{72}}, 09690 (2009).


\bibitem{other}  R. R. Caldwell \& M. Kamionkowski, Ann.\ Rev.\ Nucl.\ Part.\ Sci. {\bf{59}}, 397 (2009).

\bibitem{mad} M.~A.~Dantas, J.~S.~Alcaniz, D.~Mania and B.~Ratra,  Phys.\ Lett.\  B {\bf 699}, 239 (2011). arXiv:1010.0995 [astro-ph.CO]

\bibitem{quint} J.~A.~Frieman {\it et al.}, Phys.\ Rev.\ Lett.\  {\bf 75}, 2077 (1995); R. R. Caldwell, R. Dave and P. J. Steinhardt, Phys. Rev. Lett. {\bf{80}}, 1582 (1998);  R.~R.~Caldwell and E.~V.~Linder, Phys.\ Rev.\ Lett.\  {\bf 95}, 141301 (2005); F.~C.~Carvalho {\it et al.}, Phys.\ Rev.\ Lett.\  {\bf 97}, 081301 (2006). astro-ph/0608439.

\bibitem{para} Y. Wang and P. M. Garnavich, Astrophys. J. {\bf 552}, 445 (2001); C. R. Watson and R. J. Scherrer, Phys. Rev. D {\bf 68}, 123524 (2003); P.S. Corasaniti et al., Phys. Rev. D {\bf 70}, 083006 (2004); V. B. Johri, astro-ph/0409161;  H. K. Jassal, J. S. Bagla, and T. Padmanabhan, Mon. Not. Roy. Astron. Soc. {\bf 356}, L11 (2005); Y. Wang, Phys. Rev. {\bf{D 77}}, 123525 (2008).

\bibitem{para1} Y. Wang and M. Tegmark, Phys. Rev Lett. {\bf 92}, 241302 (2004).

\bibitem{generalized}E. M. Barboza Jr. {\it et al.}, Phys. Rev. {\bf{D80}} {{043521}} (2009). arXiv:0905.4052 [astro-ph.CO]

\bibitem{linear} A. R. Cooray and D. Huterer, Astrophys. J. {\bf 513}, L95 (1999); P. Astier, Phys. Lett. B, {\bf 500}, 8 (2001); M. Goliath {\it et al.}, Astron. Astrophys. {\bf{380}}, 6 (2001); J. Weller and A. Albrecht, Phys. Rev D {\bf 65}, 103512 (2002).

\bibitem{cpl} M. Chevallier and D. Polarski, Int. J. Mod. Phys. D {\bf 10}, 213 (2001); E. V. Linder, Phys. Rev. Lett. {\bf 90}, 091301 (2003).

\bibitem{efs} G. Efstathiou, Mon. Not. Roy. Astron. Soc., {\bf 310}, 842 (1999).

\bibitem{edesio} E.~M.~Barboza Jr. and J.~S.~Alcaniz, Phys.\ Lett.\  B {\bf 666}, 415 (2008).  arXiv:0805.1713 [astro-ph]

\bibitem{pavon} G.~Izquierdo and D.~Pavon,  Phys.\ Lett.\  B {\bf 688}, 115 (2010).

\bibitem{method}  Z.~K.~Guo, N.~Ohta and Y.~Z.~Zhang,  Phys.\ Rev.\  D {\bf 72}, 023504 (2005).

\bibitem{union} R.~Amanullah {\it et al.},  Astrophys.\ J.\  {\bf 716}, 712 (2010).

\bibitem{sdss} R. Kessler {\it et al.}, Astrophys. J. Suppl. Ser. {\bf{185}}, 32 (2009).

\bibitem{salt2} J. Guy {\it et al.}, Astron. Astrophys. {\bf{466}}, 11 (2007).

\bibitem{mlcs2k2} M. M. Phillips, Astrophys. J. {\bf{413}}, L105 (1993); A. G. Riess, W. H. Press, and R. P. Kirshner, Astrophys. J. {\bf{438}}, L17 (1995); S. Jha, A. G. Riess, and R. P. Kirshner, Astrophys. J. {\bf{659}}, 122 (2007).

\bibitem{percival} W. J. Percival et al., MNRAS, {\bf{401}}, 2148 (2010).

\bibitem{bao} D. J. Eisenstein {\it et al.}, Astrophys. J. {\bf{633}}, 560 (2005).

\bibitem{blake} C.~Blake {\it et al.}, ``The WiggleZ Dark Energy Survey: testing the cosmological model with baryon acoustic oscillations at $z=0.6$,''  arXiv:1105.2862 [astro-ph.CO].

\bibitem{wmap} E. Komatsu et al.,  Astrophys. J. Suppl. Ser. {\bf{192}}, 18 (2011). 


\bibitem{phantom}R. R. Caldwell, Phys. Lett. B {\bf 545}, 23 (2002); S. M. Carroll, M. Hoffman and M. Trodden, Phys. Rev. D {\bf 68}, 023509 (2003); J.~S.~Alcaniz,  Phys.\ Rev.\  D {\bf 69}, 083521 (2004). astro-ph/0312424.

\bibitem{divideline} L.~P.~Chimento, R.~Lazkoz, R.~Maartens and I.~Quiros,  JCAP {\bf 0609}, 004 (2006);   S.~Nesseris and L.~Perivolaropoulos,  JCAP {\bf 0701}, 018 (2007).


\end{thebibliography}
\end{document}